\documentclass[final,5p,times,twocolumn]{elsarticle}

\usepackage{amssymb}

\usepackage{graphicx}

\usepackage{url}

\journal{Nuclear Instruments and Methods in Physics Research, Section A}

\begin{document}

\begin{frontmatter}



\title{Upgrade of ASACUSA's Antihydrogen Detector}


\author[1]{V. Kraxberger\corref{cor1}}
\ead{viktoria.kraxberger@oeaw.ac.at}
\author[1]{C. Amsler}
\author[2]{H. Breuker}
\author[1]{S. Chesnevskaya}
\author[3,4]{G. Costantini}
\author[5,6]{R. Ferragut}
\author[6]{M. Giammarchi}
\author[1]{A. Gligorova}
\author[3,4]{G. Gosta}
\author[7]{H. Higaki}
\author[1]{E. D. Hunter}
\author[1]{C. Killian}
\author[1]{V. Kletzl\fnref{vici}}
\author[8]{N. Kuroda}
\author[1,9]{A. Lanz}
\author[3,4]{M. Leali}
\author[1]{V. Mäckel}
\author[6,10]{G.Maero}
\author[11]{C. Malbrunot\fnref{chloe}}
\author[3,4]{V. Mascagna}
\author[8]{Y. Matsuda}
\author[3,4]{S. Migliorati}
\author[1]{D. J. Murtagh}
\author[12]{Y. Nagata}
\author[1,9]{A. Nanda}
\author[11,9]{L. Nowak}
\author[6,10]{E. Pasino}
\author[6,10]{M. Romé}
\author[1]{M. C. Simon}
\author[13]{M. Tajima}
\author[5,6]{V. Toso}
\author[2]{S. Ulmer}
\author[3,4]{L. Venturelli}
\author[1,9]{A. Weiser}
\author[1]{E. Widmann}
\author[11]{T. Wolz}
\author[2]{Y. Yamazaki}
\author[1]{J. Zmeskal}
	
\address[1]{Stefan-Meyer-Institute for Subatomic Physics, Austrian Academy of Science, Vienna, Austria}
\address[2]{Ulmer Fundamental Symmetries Laboratory, RIKEN, Saitama, Japan}
\address[3]{Dipartimento di Ingegneria dell’Informazione, Università degli Studi di Brescia, Brescia, Italy}
\address[4]{INFN sez. Pavia, Pavia, Italy}
\address[5]{Politecnico di Milano, Milan, Italy}
\address[6]{INFN sez. Milano, Milan, Italy}
\address[7]{Graduate School of Advanced Science and Engineering, Hiroshima University, Hiroshima, Japan}
\address[8]{Institute of Physics, Graduate School of Arts and Sciences, University of Tokyo, Tokyo, Japan}
\address[9]{University of Vienna, Vienna Doctoral School in Physics, Vienna, Austria}
\address[10]{Dipartimento di Fisica, Università degli Studi di Milano}
\address[11]{Experimental Physics Department, CERN, Geneva, Switzerland}
\address[12]{Department of Physics, Tokyo University of Science, Tokyo, Japan}
\address[13]{RIKEN Nishina Center for Accelerator-Based Science, Saitama, Japan}

\cortext[cor1]{Corresponding author at Stefan-Meyer-Institute for Subatomic Physics of the Austrian Academy of Science, Kegelgasse 27, 1030 Vienna, Austria}

\fntext[vici]{Present address: Paul Scherrer Institut, Villigen, Switzerland}
\fntext[chloe]{Present address: TRIUMF, Vancouver, Canada}


\begin{abstract}
The goal of the ASACUSA (Atomic Spectroscopy And Collisions Using Slow Antiprotons) CUSP experiment at CERN's Antiproton Decelerator is to measure the ground state hyperfine splitting of antihydrogen in order to test whether CPT invariance is broken.

The ASACUSA hodoscope is a detector consisting of two layers of 32 plastic scintillator bars  individually read out by two serially connected silicon photomultipliers (SiPMs) on each end. Two additional layers for position resolution along the beam axis were scintillator fibres, which will now be replaced by scintillating tiles placed onto the existing bars and also read out by SiPMs.
If the antiproton of antihydrogen annihilates in the centre of the hodoscope, particles (mostly pions) are produced and travel through the various layers of the detector and produce signals.

The hodoscope was successfully used during the last data taking period at CERN. The necessary time resolution to discriminate between particles travelling through the detector from outside and particles produced in the centre of the detector was achieved by the use of waveform digitisers and software constant fraction discrimination. The disadvantage of this readout scheme was the slow readout speed, which was improved by two orders of magnitude. This was done by omitting the digitisers and replacing them with TDCs reading out the digital time-over-threshold (ToT) signal using leading edge discrimination.

\end{abstract}




\end{frontmatter}



\section{Introduction}
\label{ch:intro}
The CPT symmetry (charge conjugation, parity transformation and time reversal) is a fundamental part of the Standard Model of particle physics. If it is valid then matter and antimatter must have equal or sign-opposite intrinsic properties. 
The matter--antimatter asymmetry observed in our universe cannot be explained by the already found CP violations \cite{Zyla2020}. The ASACUSA experiment plans to create an antihydrogen beam to measure the hyperfine splitting of ground-state $\overline{\mathrm{H}}$ using Rabi spectroscopy \cite{Widmann2019, Malbrunot2018}. The hodoscope is used to analyse the number of arriving antihydrogen atoms by detecting their annihilation products  \cite{Kolbinger2021}.\\
In the previous ASACUSA CUSP beam times the rate of antihydrogen reaching the detector was only a few Hz, but for the upcoming beam times it is expected that the antihydrogen production will be more efficient and a faster readout is needed. The maximum readout rate with the previous setup was at $\approx$ 50 Hz, the goal was to reach 1 kHz. For that the previously used data acquisition (DAQ) system \cite{Sauerzopf2016, Fleck2018} had to be modified. 

\section{The Hodoscope Detector}
\label{ch:hodoscope}
\begin{figure}[h]
\centering
\includegraphics[width=0.45\textwidth]{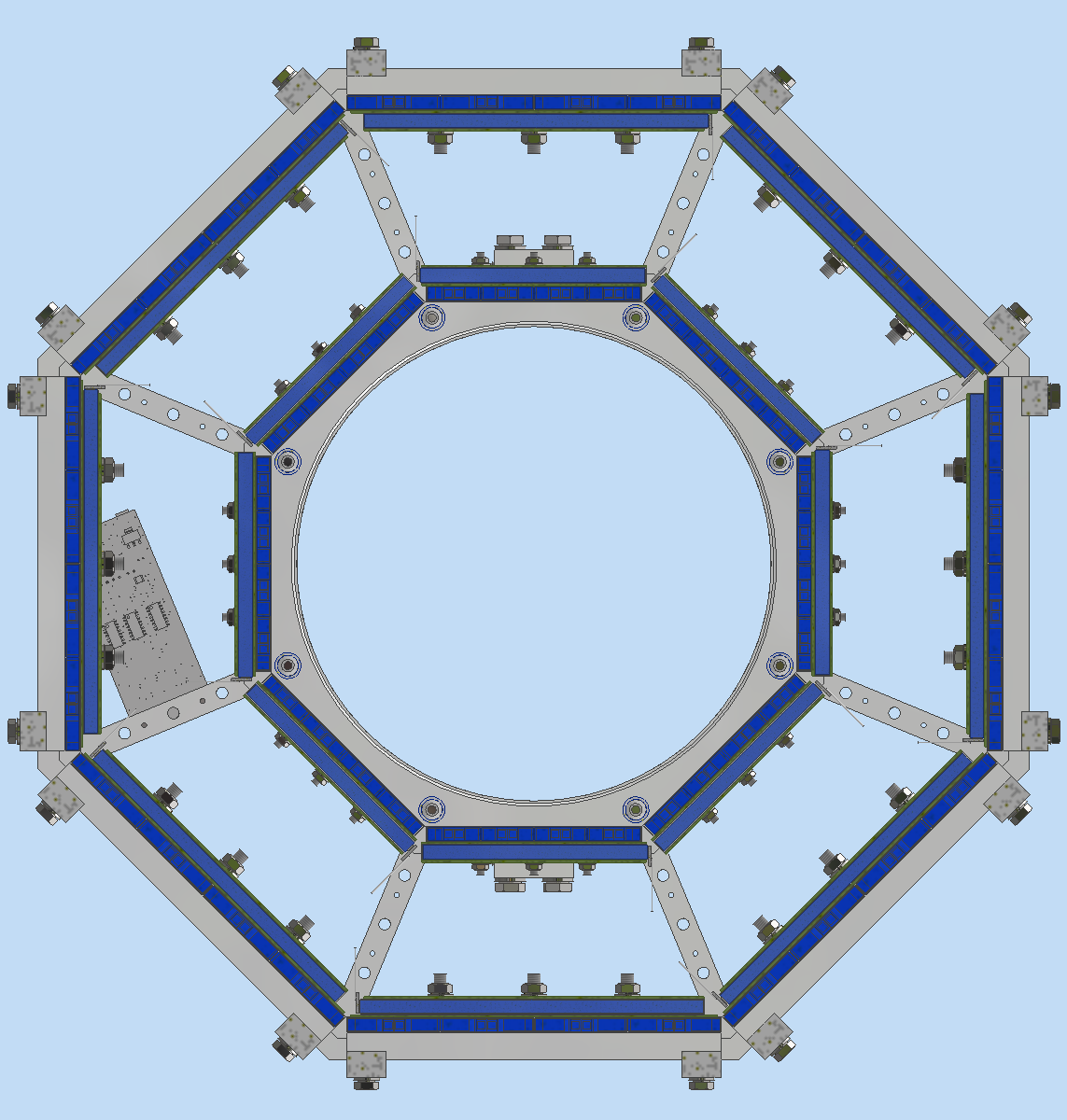}
\caption{Front view of the hodoscope with an outer and inner layer of 32 scintillator bars and 120 scintillator tiles each. The distance between two opposite outer bars is 350 mm.}
\label{fig:compare_amps}
\end{figure}

The hodoscope (Fig. \ref{fig:compare_amps}) is a two-layered octagonal barrel-type detector consisting of 32 EJ 200 scintillator bars per layer. The outer bars have a dimension of $450 \times 35 \times 5$ mm$^3$, the inners are $300 \times 20 \times 5$ mm$^3$. 
Light guides in the shape of trapezoidal prisms are attached on both ends of the bars with a length of 40 mm (inner layer) and 75 mm (outer layer). They reduce the cross section of the bars down to $8 \times 5$  mm$^2$ matching the active area of the used silicon photomultipliers.
Until recently the position resolution in beam direction was provided by scintillating fibres (100 in the outer, 64 in the inner layer) which are now being replaced by 240 EJ 200 scintillator tiles (15 per octagon side and layer). On the outer layer the $129 \times 30 \times 5$ mm$^3$ tiles will be placed perpendicularly onto each section of four bars. The same is true for the inner layer,  where the tiles have a size of $84 \times 20 \times 5$ mm$^3$. The bars are readout using two $3 \times 3$ KETEK PM3350-TS SiPMs on each end, for the fibres one $3 \times 3$ KETEK PM3350-EB was used. The new tile-readout is done with two serially connected Broadcom AFBR-S4N33C013 $3 \times 3$ SiPMs on one side. 

\section{Data Acquisition Hardware and Software Upgrades}
\label{ch:daq}
The DAQ is done using CAEN V1190A/B TDCs and CAEN V2495 FPGA units together with the MIDAS \cite{MIDAS} software. A VME controller (Struck SIS3104) connects them to a PC via optical link fibre. Both hardware and software were upgraded and reprogrammed to allow a higher readout rate. The MIDAS frontend was adapted to use a polled trigger, already allowing readout at a rate of 300 Hz and the TDC readout was changed to block transfer mode, such that now it is possible to record events at rates of over 1 kHz. Compared with the previous data acquisition, this represents an improvement by two orders of magnitude.

A trigger for the DAQ is produced using two CAEN V2495 FPGA logic modules replacing the before used combination of several NIM logic units and one VME CAEN V2495 FPGA. This trigger is produced for events where there is a signal on both ends of one scintillator bar (to exclude triggering on noise of a single SiPM) for at least one inner and one outer bar. It is then distributed to each TDC and the SIS3104.

\section{Time and Position Resolution}
\label{sec:hodo_tests}
In the new setup only digital time-over-threshold signals are used which are produced by leading edge discrimination. This introduces a jitter depending on the rise time of the signals. Due to the difference in rise times, the ToT signal is not produced at a certain fraction of the amplitude, but at a fixed threshold. If the signals are produced with SiPMs operated in saturation, the amplitude of all signals is maximal ($\approx$ 800 mV) and the rise time is minimised. When taking the leading edge (LE) times at a specific threshold voltage, the signals are therefore more constant as the rise time does not vary as much.

\subsection{Resolution with a Test Setup}
\label{subsec:test_setup}
\begin{figure}[h]
\centering
\includegraphics[width=0.45\textwidth]{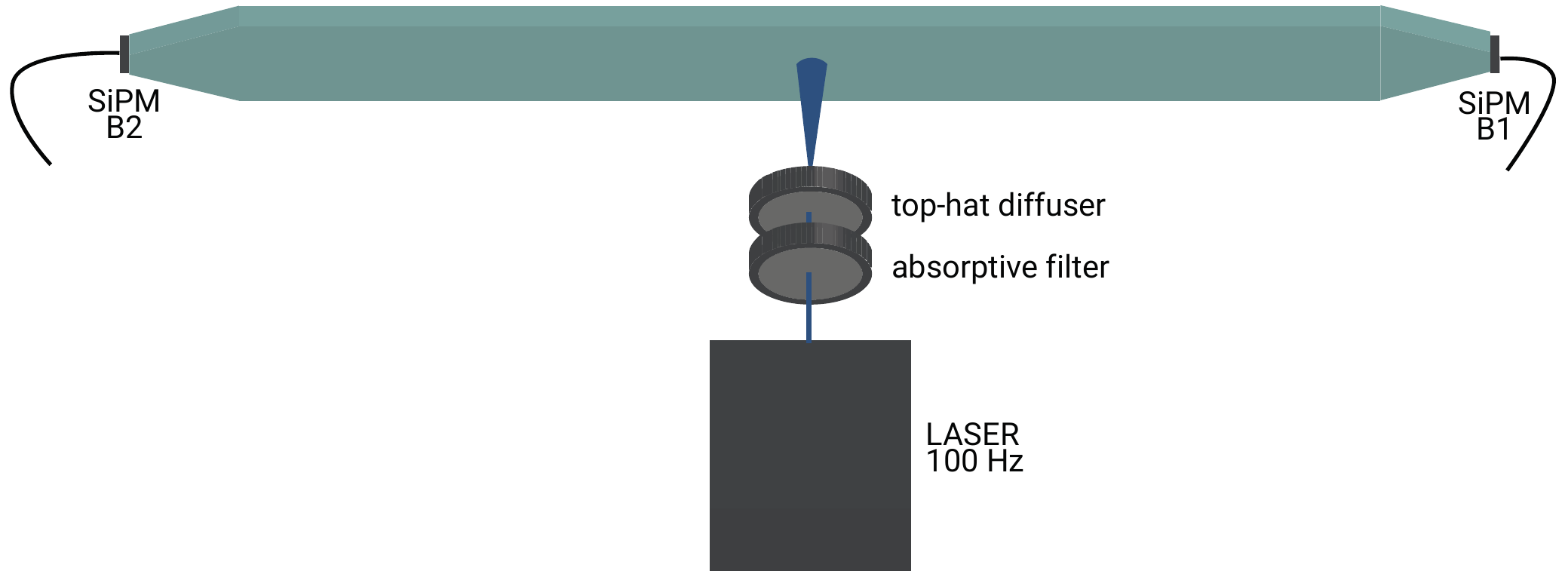}
\caption{A sketch of the setup with a pulsed laser positioned in front of the scintillator bar. The laser beam first goes through an absorptive filter to lower the intensity and then through a top-hat diffuser so the beam has a diameter of $\approx$ 5 mm.}
\label{fig:test_setup}
\end{figure}

In the test setup, see Fig. \ref{fig:test_setup}, a pico-second laser pulse was used to produce signals in the scintillator bar, read out at each end with an FPGA and TDCs similar to the hodoscope setup. The measurement was done with the laser irradiating various points along the scintillator bar. The time differences of the leading edges of the signals on both ends were fitted with a Gaussian distribution. 

\begin{figure}[h]
\centering
\includegraphics[width=0.45\textwidth]{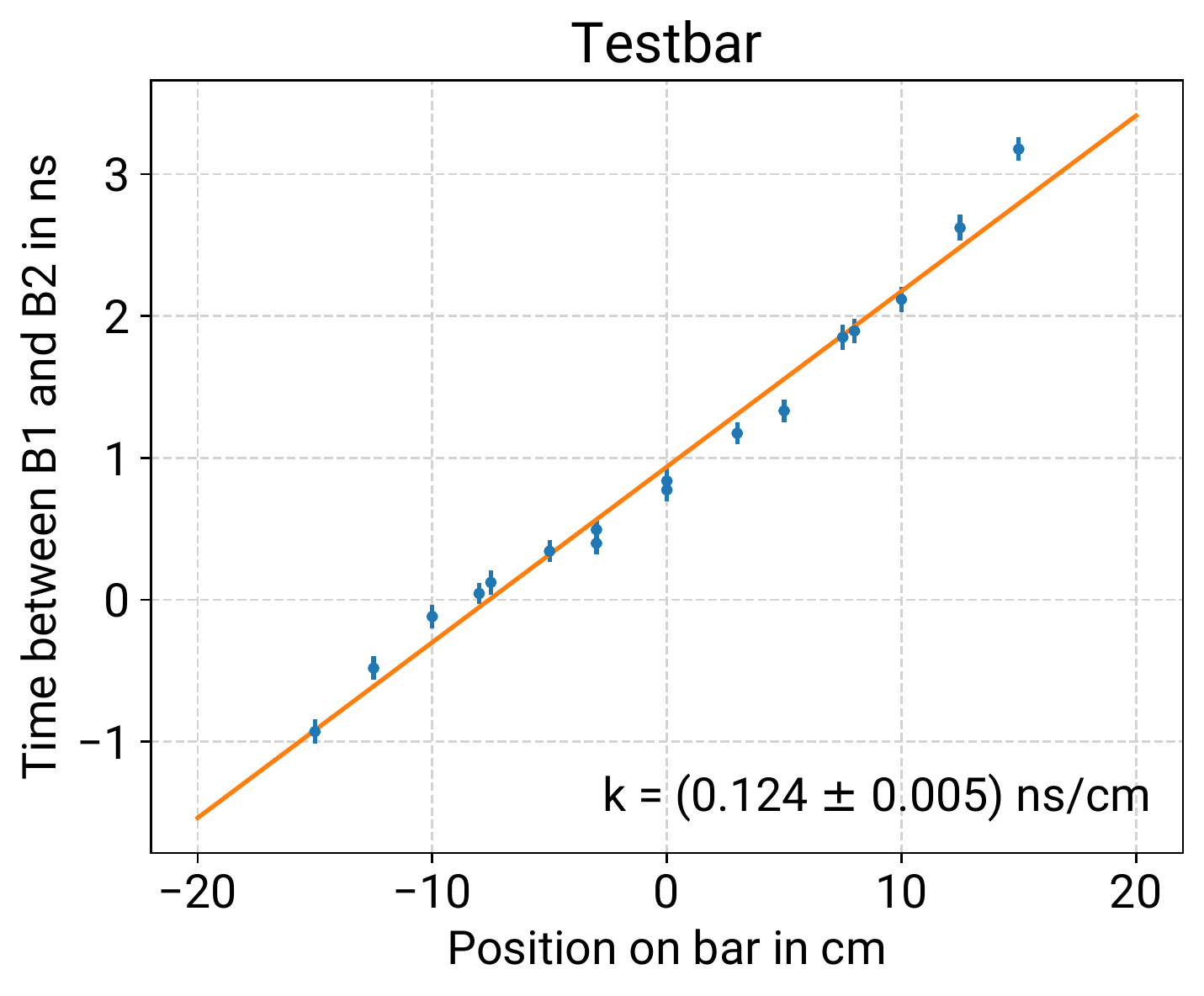}
\caption{The means of the time differences between the two ends of the bar plotted against the position of the laser along the bar, where 0 cm is the middle. The y-errorbars show the standard deviations. The orange line shows a linear fit, where the resulting slope is $k = (0.124 \pm 0.005)$ ns/cm.}
\label{fig:positions1}
\end{figure}

In Fig. \ref{fig:positions1} these time differences with the standard deviation of the Gaussian fit as y-errorbars are plotted against the positions of the laser along the bar. There is a linear dependence between time differences and positions, where the slope $k$ is indicated in the graph. Using this linear function the position resolution could also be determined as $\approx$ 1 cm standard deviation. Compared to the resolution with the previous setup this would be an improvement by a factor of 3 \cite{Kolbinger2021}.

\subsection{Hodoscope Resolution}
\label{subsec:hodo_res}
\begin{figure}[h]
\centering
\includegraphics[width=0.45\textwidth]{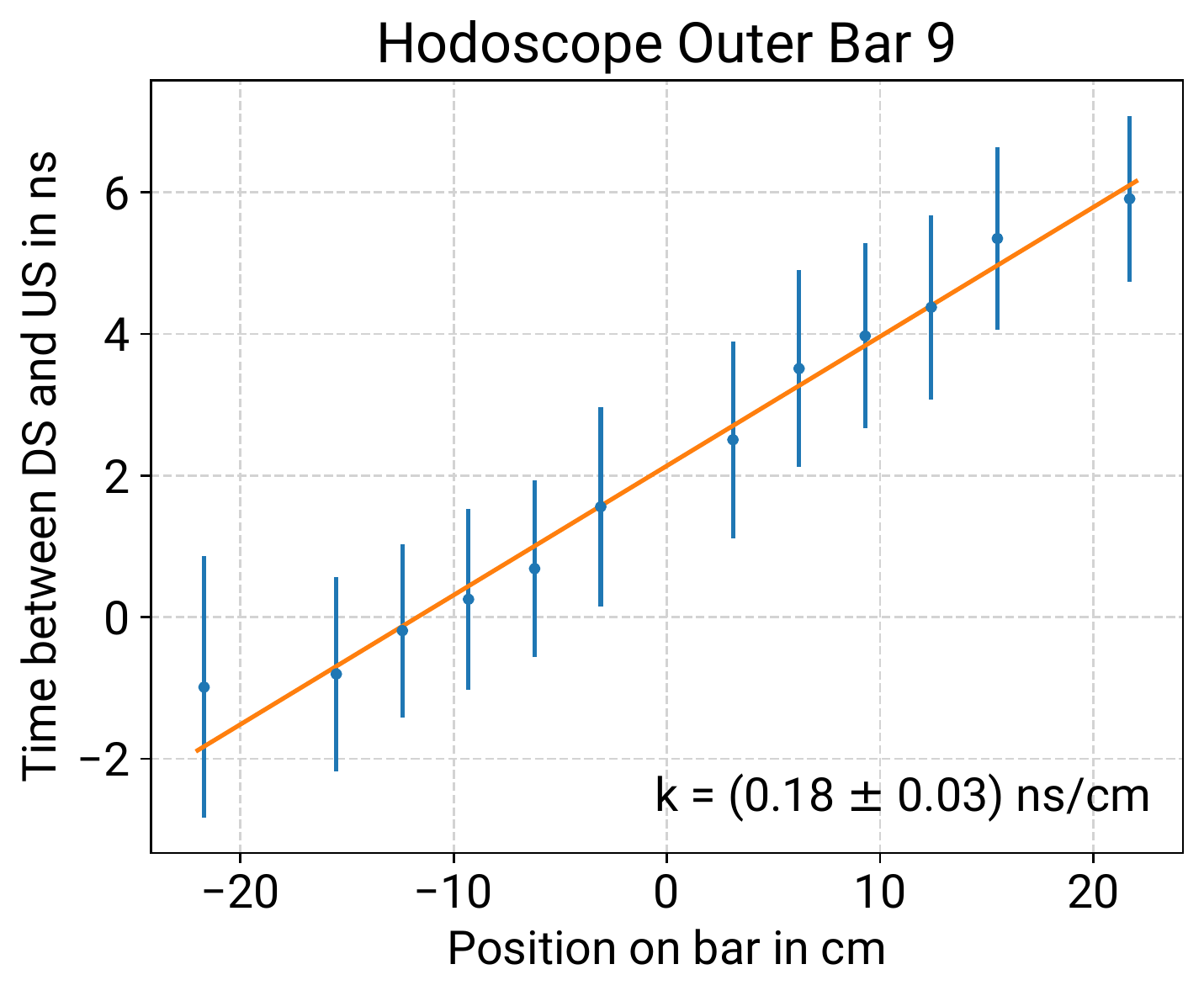}
\caption{The means of the time differences between the two ends of the bar plotted against the position of the tile along the bar. The y-errorbars show the standard deviations. The orange line shows a linear fit, where the resulting slope is $k = (0.18 \pm 0.03)$ ns/cm.}
\label{fig:positions2}
\end{figure}

To test these results with the whole hodoscope setup using cosmic rays, coincidences with a pair of fibres (one inner, one outer) were used to consider only signals travelling through the detector perpendicular to the beam axis. Thus, the position of the hit along the bars should have been known, but the results were very poor and no position resolution could be achieved. Nevertheless there were already plans to replace the fibres with scintillator tiles to improve the tracking efficiency. Onto each section of 4 scintillator bars, 15 tiles will be placed perpendicularly. \\
Looking at the time differences between downstream (DS) and upstream (US) leading edge times in coincidence with one the 15 tiles, a linear dependance can be seen, Fig. \ref{fig:positions2}. The mean of the measured standard deviations on this bar is $\langle \sigma \rangle = (1.4 \pm 0.1)$ ns. This can still be improved by increasing the amplification of the SiPM signals and therefore minimising the time walk of the leading edges.

\section{Conclusion}
\label{sec:conclusion}
We have performed changes in the DAQ hardware and software to achieve a possible readout rate of over 1 kHz in ASACUSA's antihydrogen detector. The hardware now only consists of VME modules and the signals are recorded as time-over-thresholds via leading edge discrimination. From a test setup we could conclude that a very good time resolution is possible, but this has not been achieved yet with the hodoscope. The currently performed upgrade to the use of scintillator tiles is promising for future measurements.

\section*{Acknowledgments}
This work was supported by Austrian Science Fund (FWF) [P 32468], [W1252-N27], and [P 34438]; the JSPS KAKENHI Fostering Joint International Research B 19KK0075 and Grant-in-Aid for Scientific Research B 20H01930; Special Research Projects for Basic Science of RIKEN; Università di Brescia and Istituto Nazionale di Fisica Nucleare; and the European Union’s Horizon 2020 research and innovation programme under the Marie Skłodowska-Curie grant agreement No 721559. For the purpose of open access, the author has applied a CC BY public copyright licence to any Author Accepted Manuscript version arising from this submission.



  \bibliographystyle{elsarticle-num} 
  \bibliography{citations}





\end{document}